\documentstyle [prl,aps,twocolumn,epsfig,psfig]{revtex}
\newskip\humongous \humongous=0pt plus 1000pt minus 1000pt
\def\caja{\mathsurround=0pt}
\def\eqalign#1{\,\vcenter{\openup1\jot \caja
	\ialign{\strut \hfil$\displaystyle{##}$&$
	\displaystyle{{}##}$\hfil\crcr#1\crcr}}\,}
\newif\ifdtup

\relax

\begin{document}
\title  {New String Excitations in the Two-Higgs Standard Model 
}
\bigskip

\author{C. Bachas} 

\address{Centre de Physique Th\'eorique, Ecole Polytechnique, 91128 Palaiseau;
France}

\author{B. Rai and T.N. Tomaras }

\address{Department of Physics and Institute of Plasma Physics,
University of Crete and FO.R.T.H., \\ 
 P.O.Box 2208, 710 03 Heraklion, Crete; Greece}
\date{\today}
\maketitle

\begin{abstract}
We establish the existence of a static, classically stable string solution 
in a region of parameters of  the generic two-Higgs Standard Model.
In an appropriate limit of parameters, the solution
reduces to the well-known soliton of the $O(3)$
non-linear sigma model.
\end{abstract}

\narrowtext
\medskip

The effective theory of electroweak interactions may contain more
than the single Higgs doublet of the minimal Standard Model.
 Theoretical arguments in favour of an extended
Higgs sector \cite{vivlio}  include supersymmetry and  string theory, 
as well as the possibility of electroweak baryogenesis
\cite{Baryon}.   
Extra Higgs scalars  give rise to a  richer spectrum
of non-perturbative excitations, such as 
  membrane defects \cite{membranes}
and new unstable sphalerons \cite{BTT}. In  this letter we will establish 
the existence of a new string excitation in the two-Higgs standard model
(2HSM), and we will argue that it is  stable in a
region of parameter space  extending  into  weak coupling. 

A systematic way to search for such non-topological 
 excitations has been outlined
by two of us in refs. \cite{ribbons,vortices}: one considers
appropriate  limits of parameters so that field space acquires
non-trivial topology, thus allowing for topologically-stable solitons.
These solitons will  generally  continue to exist when the parameters
are relaxed by a `small amount', so that 
the   degrees of freedom that were frozen or decoupled in the limit
affect  only slightly the dynamics.
A possible exception to this rule occurs when the limiting soliton
has zero modes other than its center-of-mass position. 
Small corrections may lift in this case the degeneracy, and  either
fix or completely destabilize the excitation. This is exemplified
by the Belavin-Polyakov soliton \cite{Belavin}, whose arbitrary
scale can be fixed by embedding it into a gauged linear
sigma model \cite{vortices}.

 The purpose of the present letter
is to show that one can embed this same soliton as a stable
static string defect in the 2HSM. The key observation
is that in an appropriate  limit of parameters the 
only relevant dynamical field is the relative SU(2) phase of the two
higgses, partially-constrained to lie on 
 a two-sphere $S^2$. 
Our string defects are characterized by the fact that the mapping 
of the transverse  two-dimensional plane onto $S^2$ has non-vanishing
winding number.  In contrast
to the  previously discussed vortex 
strings \cite{Zstring} the  defects described here carry no
net electroweak flux, have no symmetry restoration in their core,
 and do not exist in the one-Higgs standard
model.

The Lagrangian of the 2HSM is
${\cal L} = 
  -{1\over 4} W^a_{\mu\nu} W^{a\mu\nu} -{1\over 4} Y_{\mu\nu} Y^{\mu\nu}
 + \vert D_\mu H_1\vert^2 + \vert D_\mu H_2\vert^2 
 - V(H_1, H_2)$, 
where $W^a_{\mu\nu}=\partial_\mu W^a_\nu - \partial_\nu W^a_\mu - 
g \epsilon^{abc} W^b_\mu W^c_\nu$, $Y_{\mu\nu}=\partial_\mu Y_\nu - 
\partial_\nu Y_\mu$. 
The physical $Z^0$ and photon fields are
$Z_\mu=W^3_\mu {\rm cos}\theta_W - Y_\mu {\rm sin}\theta_W$ and
$A_\mu=W^3_\mu {\rm sin}\theta_W + Y_\mu {\rm cos}\theta_W$
and 
${\rm tan} \theta_W = g^\prime/ g$. 
Both Higgs doublets have hypercharge equal to one, 
their covariant derivatives are defined by 
$D_\mu H_{1(2)} 
=\Bigl(\partial_\mu + {i\over2} g \tau^a W^a_\mu + {i\over2} g^\prime 
Y_\mu \Bigr) H_{1(2)}$, 
while their potential reads
\begin{equation}
\eqalign{
&V(H_1, H_2)  =  \lambda_1 \Bigl( \vert H_1\vert^2 - {v_1^2\over2}\Bigr)^2 +
\lambda_2 \Bigl( \vert H_2\vert^2 - {v_2^2\over 2}\Bigr)^2  \cr
& + \lambda_3 \Bigl( \vert H_1\vert^2  + \vert H_2\vert^2 - 
{v_1^2+v_2^2\over2}\Bigr)^2 
+ \lambda_4 \Bigl[ \vert H_1\vert^2 \vert H_2\vert^2  \cr & 
- (H_1^\dagger
H_2)  (H_2^\dagger H_1)\Bigr] 
+ \lambda_5 \Bigl[{\rm Re }
H_1^\dagger H_2 -{v_1 v_2\over2} cos\xi\Bigr] ^2 \cr &\ \ \ \ \ \ \ \   + 
\lambda_6 \Bigl[ {\rm Im}H_1^\dagger H_2-{v_1 v_2\over2}
sin\xi\Bigr]^2\ .
\label{potential}
}
\end{equation}
 This is the most general potential \cite{vivlio}
subject to the condition that
both CP invariance and a discrete $Z_2$ symmetry ($H_1\rightarrow -H_1$)
are only broken softly. The softly broken $Z_2$ symmetry is there
to suppress unacceptably large 
flavor-changing neutral 
currents.                          

Assuming all $\lambda_i$ are
positive, the minimum of the 
potential is, up to  gauge transformations, at
$<H_1>=(0, v_1/\sqrt{2})$, $<H_2>=e^{i\xi}\; (0, v_2/\sqrt{2})$.
The perturbative spectrum consists of the electroweak gauge bosons
with masses $m_W^2=g^2(v_1^2+v_2^2)/4$ and $m_Z=m_W/cos\theta_W$ ,  
a charged Higgs scalar  $H^+$ with
mass $m_{H^+}^2=\lambda_4(v_1^2 + v_2^2)/2$, and three
neutral scalars, one CP-odd ($A^0$)  and two CP-even ($h^0$ and $H^0$) , 
whose masses depend on the value of $\xi$.
The neutral mass matrix simplifies considerably for $\sin\xi \cos\xi=0$.
For $\xi=\pi/2 $
the neutral masses are
$m_{A^0}^2=\lambda_5(v_1^2+v_2^2)/2$, 
$m_{h^0}^2=\zeta-\sqrt{\eta}$ and 
$m_{H^0}^2=\zeta+\sqrt{\eta}$,  where 
$\zeta \equiv (\lambda_1 v_1^2+\lambda_2 v_2^2)
+(\lambda_3+{1\over 4} \lambda_6)(v_1^2+v_2^2)$ and 
$\eta \equiv (\lambda_1 + \lambda_3 
-{1\over 4} \lambda_6)^2 v_1^4 
+(\lambda_2+\lambda_3-{1\over 4} \lambda_6)^2 v_2^4
+ 2 v_1^2 v_2^2 \Bigl((\lambda_3+{1\over 4} \lambda_6)^2+\lambda_3
\lambda_6-(\lambda_3-{1\over 4} \lambda_6)
(\lambda_1+\lambda_2)- \lambda_1\lambda_2 \Bigr)$.
For $\xi=0$ the masses of $A^0$, $h^0$ and $H^0$ 
are given by the same expressions with 
$\lambda_5$ and $\lambda_6$ interchanged. 
To simplify the discussion we  take $\xi=\pi/2$, 
$g^\prime = \lambda_3 = \lambda_4 = 0$, and   drop the decoupled
$U(1)$ gauge field $Y^\mu$. We will comment on these assumptions at
the end.

Since we  are interested in classically-stable static string
solutions, we  choose the $W_0^a = 0$ gauge and 
work with static field configurations, so that 
Gauss' law is automatically satisfied. 
Furthermore, for strings stretching in the $x_3$ direction we  take all fields 
to be independent of $x_3$ and  put $W_3^a=0$. It is easy to verify that 
stable minima of the ensuing two-dimensional energy functional
correspond to stable infinite-string excitations in the original model.
Unlike  the Nielsen-Olesen-type  vortex strings, the
 finite-energy solutions of
interest to us will have $\vert H_{1(2)} \vert \neq 0 $ everywhere. 
This fact allows us to use the radial
representation of the two doublets, and write them   
as $H_{1(2)}=F_{1(2)} U_{1(2)} (0\, , \,1) $. 
$F_{1(2)}$ are two positive functions, while $U_{1(2)}$ 
are two smooth $SU(2)$
valued functions on the plane. The space of smooth maps 
from the two-dimensional
plane into $SU(2)\equiv S^3$ is topologically trivial, and there is no
topological obstruction in using the remaining 
freedom of smooth transverse-space-dependent
gauge transformations to set either $U_1$ or $U_2$ 
equal to the identity matrix. We will choose 
$U_1=1$ so that  the most general Higgs configuration takes the form
\begin{equation}
H_1 = F_1 \left( \matrix{ 0 \cr  1 \cr} \right)
\ \ {\rm and} \ \ 
H_2 = F_2 \,{\rm exp}(- i\Theta{\bf n}\cdot{\bf \tau})
\left(\matrix{0 \cr 1\cr}\right)
\label{spherical} 
\end{equation}
with ${\bf n}\cdot{\bf n}=1$ 
and ${\bf \tau}$ the Pauli matrices.

In the naive limit  
$\lambda_1, \lambda_2, \lambda_5 \to \infty$ the 
$F_1$, $F_2$ and $\Theta$ freeze at their vacuum values
$v_1/\sqrt{2}$, $v_2/\sqrt{2}$ and $\xi=\pi/2 $ respectively. This restricts 
the Higgs
target space to a two-sphere $S^2$, parametrized by the unit vector
${\bf n}$. Sending  $\lambda_6 \to 0$ removes the potential of ${\bf n}$.
In order to decouple the gauge field we take  furthermore
the limit
$g\to 0$. This  is a priori dangerous because it renders the gauge fixing 
(\ref{spherical})  singular.  To remedy the situation
 we  must also  take  
$v_1\to\infty$  while keeping the gauge-boson mass $m_W\sim g v_1$ finite. 
We will see explicitly later on that a finite gauge-boson mass is
indeed necessary for our stability analysis. 

 In the above limit  of parameters one is left
 with  an
effective O(3) non-linear $\sigma$-model  ${\cal L} \sim
 v_2^2 (\partial_\mu {\bf n})^2$  describing  the
dynamics of the unit-vector field ${\bf n}(x)$.
The $\sigma$-model
has well-known topological 
solitons in two spatial dimensions \cite{Belavin}, characterized
by an integer winding number $N$. This is the number of times ${\bf n}(x)$
winds around $S^2$ as its argument covers the plane transverse
to the defect. 
It is convenient to 
parametrize $S^2$ through a stereographic projection:  
$n^1+i n^2 = 2 \Omega / (1+|\Omega|^2), 
n^3 = (1-|\Omega|^2) / (1+|\Omega|^2)$.
The $N=1$ string, extending along the $x_3$-axis, and  
 with the boundary condition
$n^3\to 1$ at $\infty$  is then given by
\begin{equation}
\Omega = \rho e^{i\alpha} /(z-z_0)
\label{BP2}
\end{equation}
where  $z \equiv x_1+i x_2$. The position $z_0$, scale 
$\rho$ and angle $\alpha$ are arbitrary parameters
corresponding  to soliton zero modes. The value $n^3=1$ at $\infty$
will be imposed
 by the requirement of finiteness of energy once,
as we do below, we let 
$\lambda_6 \neq 0$.

Following the same steps as in ref. \cite{vortices}, we
want now to relax slightly the above limits on the parameters, and study
the fate of the solution  (\ref{BP2}). A simple scaling argument
shows that the  potential terms  tend 
to shrink the soliton to zero size, while the gauge interactions tend 
to blow it up.  Our task will be to
show that in a region of parameter space the soliton is stabilized
at some fixed scale  $\bar\rho$. 
Our semiclassical analysis looks at first sight incompatible with
the strong scalar-coupling limit considered above. To resolve this apparent
contradiction, we must  redefine the limiting theory in terms of 
classically-relevant
parameters as in \cite{ribbons,vortices}. To this end, we 
rescale the Higgs and gauge fields by 
$m_W/\sqrt{2 \lambda_1}$, and the transverse space
coordinates $x_{j}$  $(j=1,2)$ by $1/m_W$. The
energy per unit string length takes  
the form 
\begin{eqnarray}
{\cal E} & = & {m_W^2 \over 2\lambda_1} \int d^2x\; \Biggl[ 
               F_2^{\; 2} \Bigl( \sin^2\Theta (\partial_i {\bf n})^2  
               + (\partial_i \Theta)^2 \Bigr) \nonumber\\
         & + & (\partial_i F_1)^2 + (\partial_i F_2)^2 
               + {1\over 2} (F_1^2-{\tilde v}_1^2)^2 
               + {\tilde\lambda_2 \over 2} ( F_2^{\; 2} -{\tilde v}_2^2)^2 
	       \nonumber\\
         & + & {{\tilde\lambda}_5 \over 2}  F_1^{\; 2} F_2^{\; 2} \cos^2\Theta 
               + {{\tilde\lambda}_6 \over 2} ( F_1 F_2 n^3 \sin\Theta - 
               {\tilde v}_1 {\tilde v}_2)^2   \nonumber
 \\
         & + & {1\over 4} W^a_{ij} W^a_{ij} 
               + {1\over 4} {\tilde g}^2 (F_1^{\; 2} + F_2^{\; 2})
                 W^a_i W^a_i  
               - \tilde g F_2^{\; 2} J^a_i W^a_i\Biggr] \label{energy}
\end{eqnarray}
We have defined $\tilde g=g/\sqrt{2\lambda_1}$, $\tilde\lambda_m=
\lambda_m/\lambda_1$ for $m=2,5,6$,  
$\tilde v_1 \equiv \sqrt2 \cos \beta /\tilde g$,
$\tilde v_2 \equiv \sqrt2 \sin \beta /\tilde g$, 
tan$\beta=v_2/v_1={\tilde v}_2/{\tilde v}_1$,
and the current
$J^a_i=n^a \partial_i \Theta+  
\sin^2\Theta \epsilon^{abc} n^b \partial_i n^c+
\sin\Theta \cos\Theta \partial_i n^a $.
We have kept the same notation for the
rescaled coordinates and fields, as well as for the field strength
$W^a_{ij}$ which is 
defined as previously but with the gauge coupling replaced by
${\tilde g}$.  

 Since we will  keep $m_W$ different from zero
in the limiting theory, we  can  use it to fix the length and mass
scales by setting from now on $m_W =1$.  We may, furthermore, treat 
$\lambda_1$ as the semi-classical expansion parameter, which we will
take  sufficiently small so as to justify our semiclassical treatment. 
This leaves us with six
 classically-relevant parameters:
  $\tilde v_1$, $\tilde v_2$,
$\tilde \lambda_2$, $\tilde \lambda_5$, $\tilde \lambda_6$ and $\tilde g$.
The limiting theory of interest can now be defined more precisely as
follows:
\begin{equation}
\tilde v_1, \tilde \lambda_2 \to \infty; \;   
 \tilde v_2  ,   \tilde \lambda_5 \not= 0\ \ 
{\rm  fixed};  
   \;\tilde \lambda_6 \tilde v_1^2 ,\tilde g \sim  \beta \;
\to 0.
\label{limit3} 
\end{equation}
This looks different from our original naive limit, but has the same 
dynamical effects.  The first set of  conditions freeze  indeed $F_1$,
 $F_2$ and  $\Theta$  to their vacuum values, so that
the only dynamically-accessible Higgs 
 degree of freedom is  the field 
 ${\bf n}(x)$,  which parametrizes a two-sphere of non-zero radius.
The second set of conditions decouples the gauge field, and ensures that
the  dynamics of $\bf n$ 
 is described by the usual $O(3)$   
non-linear $\sigma$-model,  without any additional potential term. The
energy functional
\begin{equation}
{\cal E}_{(0)} =  {{\tilde v_2}^2 \over 2 {\lambda_1}}
             \int d^2x \;   (\partial_i {\bf n} )^2 \ ,  
\end{equation}
in this limit, admits the Belavin-Polyakov soliton
as  a (marginally-stable) solution. 

   Relaxing slightly the limit (\ref{limit3}) 
   will introduce a potential,
 unfreeze  the ``heavy
Higgs modes''  $F_1- \tilde v_1 \equiv f_1$, $F_2- \tilde v_2 \equiv f_2$
and $\Theta -\pi/2 \equiv \theta$, and couple  weakly
  the gauge field  $W_i^a$ 
to the ``light Higgs mode''  $\bf n$. These effects can be  summarized by a 
classical (non-local) effective-energy functional \cite{vortices}, 
\begin{eqnarray}
{\cal E}_{\rm eff}  & = & {\cal E}_{(0)} +  
  {{\tilde v_2}^2 \over 4 \lambda_1 }
            \;   \int d^2x \;  \Biggl[ \; 
  {\tilde \lambda_6} {\tilde v_1}^2 (n^3-1)^2
              - { (\partial_i {\bf n}\cdot\partial_i {\bf n})^2  
               \over {\tilde \lambda_2} {\tilde v_2}^2  } 
         \nonumber\\
         &   & +  {\tilde g}^2 {\tilde v_2}^2 \int d^2 y\;      
               j^a_k(x) G_{ki}(x-y) j^a_i(y)\; \Biggr] + ...  \ , 
\label{EE}
\end{eqnarray}
where  $j^a_k\equiv \epsilon^{abc}n^b\partial_k n^c$, and
 $G_{ki}$ is the two-dimensional massive
Green function satisfying
$(\delta_{jk} \Delta - \partial_j \partial_k - \delta_{jk}) G_{ki}
= \delta (\vec x - \vec y) \delta_{ji}$, and   given by  
\begin{equation}  
G_{ki} = - {1\over {(2\pi)^2}}\int d^2p
{{\delta_{ki} + p_kp_i}\over {\vec p^2 + 1}}
e^{i \vec  p\cdot (\vec  x - \vec  y)}\ .
\end{equation} 
The first correction in (\ref{EE}) is a potential term. The
second and the third come from the exchange of a heavy (radial) 
mode $f_2$ and of a vector boson, respectively.
All three are small
 compared to ${\cal E}_{(0)}$, at least  in the range of scales  $\rho \sim 1$ 
that will interest us.

  A series of comments on the above energy functional are here in order.
First, ${\cal E}_{\rm eff}$
  contains an infinite series of terms which come from
integrating out classically  the heavy Higgs modes and the vector bosons.
Inspection of  (\ref{energy}) shows that all higher   tree diagrams
contain extra heavy propagators and/or extra powers of weak couplings.
We can ensure that such corrections are indeed subleading by insisting,
for instance, that  all three heavy Higgs masses be  comparable.
 Secondly, had we set  $m_W=0$, 
 the gauge-boson exchange diagram would have been 
 IR divergent for configurations 
 with a simple power decay at infinity  like (\ref{BP2}).    
A consistent stability analysis would thus be impossible 
in this case. Finally we should point out that
Skyrme-like and potential  terms have been used to stabilize the
Belavin-Polyakov soliton in the past \cite{Z}. 
The difference  is
that in our case 
the non-local and non-renormalizable effective action (\ref{EE}) was
derived from the {\it classical} field equations of the
(renormalizable) 2HSM.

  Following \cite{vortices},  we can look for a stable solution  of
(\ref{EE}) by minimizing the correction terms with respect to the
collective coordinates of the zeroth-order solution (\ref{BP2}). 
Since ${\cal E}_{\rm eff}$ is independent of the center-of-mass position
and of the U(1) angle  $\alpha$, only the size  $\rho$ of the soliton
is important. The current $j^a_i$ 
for the Belavin-Polyakov soliton  is  
$j^a_i = \epsilon_{ij} \partial_j \phi^a$,
where $ \phi^1+i\phi^2 = 2\rho (x_1 -i x_2) / (r^2+\rho^2)$  and 
$ \phi^3 = - 2 \rho^2 / (r^2+\rho^2)$. A 
 straightforward calculation then gives:
\begin{equation}
{{\delta {\cal E}_{\rm eff}(\rho)}\over c} = 
a\rho^2- {b\over{\rho^2}} - 
\int_0^\infty dz {{z^3 ({K_1^2(z) +
 K_0^2(z)})}\over{(z^2/\rho^2)+1}}
\label{ro}
\end{equation}
where $K_0, K_1$ are the modified Bessel functions, $\delta {\cal E}_{\rm eff}
= {\cal E}_{\rm eff} -{\cal E}_{(0)}$, and we have defined the constants \\
$a \equiv {\tilde\lambda_6 \tilde v_1^2} / {2 \tilde g^2 \tilde v_2^2} \;,\; 
b\equiv 8 /{3 \tilde\lambda_2 \tilde g^2 \tilde v_2^4} \;,
\;c\equiv 2\pi \tilde g^2 \tilde v_2^4 / \lambda_1 $. \\
The shape of  $\delta {\cal E}_{\rm eff}(\rho)$,  for different values of
the parameters $a$ and $b$, was analyzed numerically. The results
are given  in Figure 1. At every point below the thick curve 
the energy has a local minimum, corresponding to a classically-stable
soliton  of size $\rho = \bar\rho (a,b)$. 
\begin{figure}[hbt] 
\centerline{\psfig{figure = 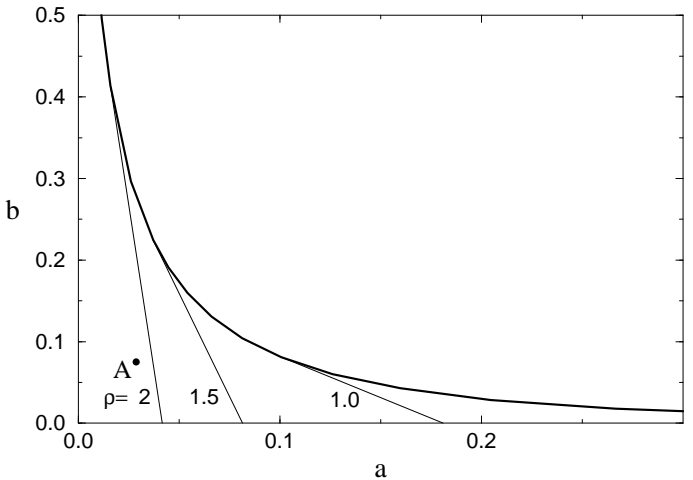, height = 72 mm}}
\caption{ The region of stability of winding string solitons.}
\end{figure}
The soliton size  stays constant
along the thin straight lines, as shown  in the figure in units of $1/m_W$.  
The assumptions that justified our perturbative treatment, and in 
particular the condition ${\bar\rho}\sim 1$, are satisfied
in a large part of the stability region. 
The validity of our analysis was confirmed independently by a
numerical minimization of the energy functional
of the 2HSM.
We use, in an obvious notation, the most general
axially symmetric ansatz 
\begin{eqnarray}
H_1 & = &\left(  \matrix{ 0 \cr  h \cr} \right), \;
H_2 = i
\left(\matrix{e^{i\phi}f\cr h_1+ih_2\cr} \right), \;
{\bf W}^3 = W^3_r {\bf e}_r + W^3_\phi {\bf e}_\phi 
\nonumber \\ 
{\bf W}^+ & = &{1\over\sqrt{2}} e^{i\phi}
\bigl((W_{r 1}+iW_{r 2}) {\bf e}_r + 
(W_{\phi 1}+iW_{\phi 2}) {\bf e}_\phi \bigr) 
\end{eqnarray}
where ${\bf W}^+ \equiv ({\bf W}^1 - i {\bf W}^2)/\sqrt{2}$.
The $\phi$-dependence is shown explicitly and the ten 
real unknown functions depend only on $r$.

Figure 2 shows the profile of the solution for 
$g=0.2$, $v_2/v_1=0.3$, $\lambda_1=0.5$, $\lambda_2=20.0$, 
$\lambda_5=1.0$ and $\lambda_6=0.0001$ (corresponding to the point
A of Figure 1) and with tension ${\cal E}=31.24 \pi m_W^2$. We did 
not plot neither $h(r)$ which is essentially equal to 6.7728
everywhere, nor $h_2$, $W_{r 1}$,
$W_{\phi 2}$, $W^3_r$ which to this accuracy are zero.
As promised and contrary to what happens for Nielsen-Olesen strings, 
the magnitudes of the Higgs fields are everywhere non zero. Also,
$\Theta \simeq \pi/2$ at all points.
\begin{figure}[hbt] 
\centerline{\psfig{figure = 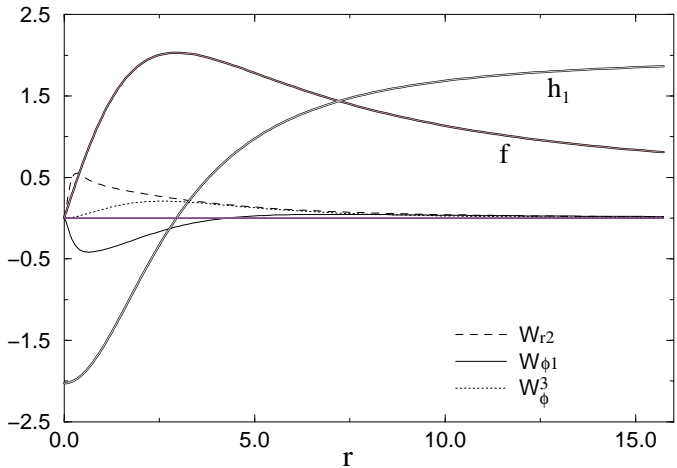, height = 72 mm}}
\caption{ The profile of a stable string.}
\end{figure}
Let us comment briefly on the parameters which were not relaxed from
zero in this discussion.  
From the analysis in \cite{vortices}  
we expect  that turning on $g^\prime$ will favour 
larger soliton radii,  without affecting 
 our qualitative  conclusions. The U(1) gauge interactions  
may even suffice by themselves  to stabilize the soliton  
against shrinking. Similarly, turning on 
$\lambda_3$ should not affect  significantly the discussion. Turning on 
 $\lambda_4$ on the other hand, to render $H^+$ massive,
 presents a technical difficulty, because
the corresponding potential term, evaluated for the zeroth-order soliton,
diverges. The divergence  is due to the slow
approach of $n^3$ to its vacuum value at large distances.
Although we don't expect this
effect to be physically-significant, our minimization procedure
would have to be modified in this case and one must check numerically
whether $H^+$ can be made heavy enough to comply with experimental
bounds. 
It is on the other hand a welcome fact of our analysis that
$\tilde \lambda_5$ need not be large for 
stable strings to exist, and that $\tilde \lambda_6$ is
favoured to be small. Realistic values of the CP violation 
parameter in the Higgs sector, which
is proportional to the difference 
$\tilde \lambda_5 - \tilde \lambda_6$
\cite{vivlio}, may be consistent with string stability 
even if $\xi$ cannot be relaxed
to very small values.
Thus, there is no a priori indication that the
 experimental limits on the parameters of the 2HSM \cite{vivlio} are
 incompatible with the existence of our defects, but this must  be
 decided ultimately by a numerical analysis 
 of the stability region \cite{tomaras}.
 The possible accelerator and/or cosmological implications of
 these solitons may of course depend sensitively on these results.

Finally, having an $S^2$ target space in the effective 
theory suggests that one should
also search for stable localized solitons
classified by the Hopf index 
$\pi_3(S^2)$ \cite{Niemi}. 
The analytic methods of this paper are not however applicable
because the zeroth order sigma model lagrangian does
not admit such Hopf solutions.

{\bf Aknowledgements}
This research was supported in part by the EU grants 
CHRX-CT94-0621 and CHRX-CT93-0340, as well as by 
the Greek General Secretariat of Research and Technology 
grant 95E$\Delta$1759. We thank the referee for
 a useful email exchange.

\vfill
                      
\eject


\begin{thebibliography}{99}

\bibitem{vivlio} J. Gunion, H. Haber, G. Kane and S. Dawson, The Higgs
Hunter's Guide, Addison-Wesley Pub. Co. (1990), and references therein.

\bibitem{Baryon} A.I. Bochkarev, S.V. Kuzmin and M.E. Shaposhnikov,
Phys. Lett. {\bf 244B}, 275 (1990) and  Phys. Rev. D{\bf 43}, 369 (1991);
N. Turok and J. Zadrozny, Phys. Rev. Lett. {\bf 65}, 2331 (1990);
L. McLerran, M. Shaposhnikov, N. Turok and M. Voloshin,
Phys. Lett. {\bf 256B}, 451 (1991).

\bibitem{membranes}  C. Bachas and T.N. Tomaras, 
Phys. Rev. Lett. {\bf 76}, 356 (1996);
G. Dvali, Z. Tavartkiladze and J. Nanobashvili, Phys. Lett. {\bf 
352B}, 214 (1995); A. Riotto and Ola Tornkvist, 
Phys. Rev. D{\bf 56}, 3917 (1997).

\bibitem{BTT} C. Bachas, P. Tinyakov and T.N. Tomaras,
Phys. Lett. {\bf B385}, 237 (1996).

\bibitem{ribbons}  C. Bachas and T.N. Tomaras,  
Nucl. Phys. {\bf B428}, 209 (1994).

\bibitem{vortices} C. Bachas and T.N. Tomaras, 
Phys. Rev. D{\bf 51}, R5356 (1995).
 
\bibitem{Belavin} A.A. Belavin and A.M. Polyakov,    
Pis'ma Zh. Eksp. Teor. Fiz. {\bf 22}, 503 (1975)
 [JETP Lett. {\bf 22}, 245 (1975)]. 

\bibitem{Zstring} 
 Y. Nambu, {\it Nucl. Phys.} {\bf B130}, 505 (1977);
  T. Vachaspati, Phys. Rev. Lett. {\bf 68}, 1977 (1992)
   and
 Nucl. Phys. {\bf B397}, 648 (1993);
 T. Vachaspati and M. Barriola, Phys. Rev. Lett. {\bf 69}, 1867 (1992);
 L. Perivolaropoulos, Phys. Lett. {\bf 316B}, 528 (1993);
 G. Dvali and G. Senjanovic, Phys. Rev. Lett. {\bf 71}, 2376 (1993);
 M. James, T. Vachaspati and L. Perivolaropoulos, 
Phys. Rev. D{\bf 46}, R5232 (1992) and Nucl. Phys. {\bf B395}, 534 (1993);
M. Earnshaw and M. James, Phys. Rev. D{\bf 48}, 5818 (1993).

\bibitem{Z}
R.A. Leese, R. Peyrard and W.J. Zakrzewski, Nonlinearity {\bf 3}, 773 (1990).  

\bibitem{tomaras} T.N. Tomaras, in progres.

\bibitem{Niemi}H.J. de Vega, Phys. Rev. D{\bf 18}, 2945 (1977);
J. Gladikowski and M. Hellmund, Phys. Rev. D{\bf 56}, 5194 (1997);
L. Faddeev and A.J. Niemi, hep-th/9610193. 

\end{thebibliography}
\end{document}